# Evolutionary Computation Algorithms for Cryptanalysis: A Study

Poonam Garg
Information Technology and Management Dept.
Institute of Management Technology
Ghaziabad, India
pgarg@imt.edu

*Abstract*— The cryptanalysis of various cipher problems can be formulated as NP-Hard combinatorial problem. Solving such problems requires time and/or memory requirement which increases with the size of the problem. Techniques for solving combinatorial problems fall into two broad groups – exact algorithms and Evolutionary Computation algorithms. An exact algorithms guarantees that the optimal solution to the problem will be found. The *exact* algorithms like branch and bound, simplex method, brute force etc methodology is very inefficient for solving combinatorial problem because of their prohibitive complexity (time and memory requirement). The Evolutionary Computation algorithms are employed in an attempt to find an adequate solution to the problem. A Evolutionary Computation algorithm - Genetic algorithm, simulated annealing and tabu search were developed to provide a robust and efficient methodology for cryptanalysis. The aim of these techniques to find sufficient "good" solution efficiently with the characteristics of the problem, instead of the global optimum solution, and thus it also provides attractive alternative for the large scale applications. This paper focuses on the methodology of Evolutionary Computation algorithms .

## I. INTRODUCTION

Cryptology is the science of building and analyzing different encryption and decryption methods. Cryptology consists of two subfields; Cryptography & Cryptanalysis. Cryptography is the science of building new powerful and efficient encryption and decryption methods. It deals with the techniques for conveying information securely. The basic aim of cryptography is to allow the intended recipients of a message to receive the message properly while preventing eavesdroppers from understanding the message. Cryptanalysis is the science and study of method of breaking cryptographic techniques i.e. ciphers. In other words it can be described as the process of searching for flaws or oversights in the design of ciphers.

## II. METHODOLOGY

### A. Genetic Algorithms

a. The General Framework Genetic Algorithms

A Genetic algorithm (GA) is a directed random search technique, invented by Holland[1], based on mechanics of natural selection and natural genetics. Genetic algorithm can find the global optimal solution in complex multi dimensional search space. Genetic algorithms are inspired by Darwin's theory of evolution.

Solution to a problem solved by genetic algorithms uses an evolutionary process. Roughly to say, Genetic algorithm begins with a set of solutions (represented by chromosomes) called population. Solutions from one population are taken and used to form a new population. This is motivated by a hope, that the new population will be better than the old one. Solutions which are then selected to form new solutions (offspring) are selected according to their fitness - the more suitable they are the more chances they have to reproduce. This is repeated until same condition (for example number of populations or improvement of the best solution) is satisfied.

b. Outlines of the Basic Genetic Algorithm

Outlines of basic genetic algorithm are described below.

1. [Start] Creation of initial population

2. [Fitness] Evaluate the fitness f(x) of each chromosome x in the population

3.  [New population] Create a new population by repeating following steps until the new population is complete

    - [Selection] Select two parent chromosomes from a population according to their fitness (the better fitness, the bigger chance to be selected)

    - [Crossover] With a crossover probability cross over the parents to form new offspring (children). If no crossover was performed, offspring is the exact copy of parents.

    - [Mutation] With a mutation probability mutate new offspring at each locus (position in chromosome).

    - [Accepting] Place new offspring in the new population

4.  [Replace] Use new generated population for a further run of the algorithm

5.  [Test] If the end condition is satisfied, stop, and return the best solution in current population

6.  [Loop] Go to step 2

    c. Creation of initial population

    There are two ways of forming this initial population. The first method consists of using randomly produced solutions created by a random number generator. This method is preferred for problem about which no prior knowledge exists or for accessing the performance of an algorithm. The second method employs a prior knowledge about the given optimization problem. Using this knowledge, a set of requirements is obtained and solutions which satisfy those requirements are collected to form an initial population.

    d. Fitness Evaluation Function

    The fitness evaluation unit acts as an interface between GA and the optimization problem. Fitness function is first derived from the objective function and used in successive genetic operations. Fitness evaluation function might be complex or simple depending on the optimization problem at hand.

    e. Selection, Crossover and Mutation operators

    It can be seen from the genetic algorithm outlines that there are three operators, namely selection, crossover and mutation. The performance is influenced mainly by two operators namely crossover & mutation.

The Selection operator is an artificial mechanism based on natural selection (e.g. the weakest individuals die off, fittest proliferate). During the reproduction phase strings composing the population are simply copied according to their objective function values. The expected number of a given string in the new population will be proportional to its fitness. Since the fittest strings have a higher probability of appearing, future generations will hopefully become fitter and fitter. Once bad parent strings are eliminated, one automatically eliminates their offspring, the offspring of their offspring and so on.

Crossover constitutes the information exchange phrase of a GA which produces diversity and innovation within the population. Strings are mated randomly to give birth to new offspring. Each element thus created will have characteristics derived from both of its parents. There are different ways of performing a crossover operator. Some common crossover operations are one-point crossover, two-point crossover, cyclic crossover and uniform crossover. Two-point crossovers are shown in Figure 1. Two individuals are randomly selected as parents from the pool of individuals formed by selection procedure and cut at a randomly chosen point p1 and p2. the trails, which are the parts after the cutting point, are swapped and two new individuals (children) are produced.

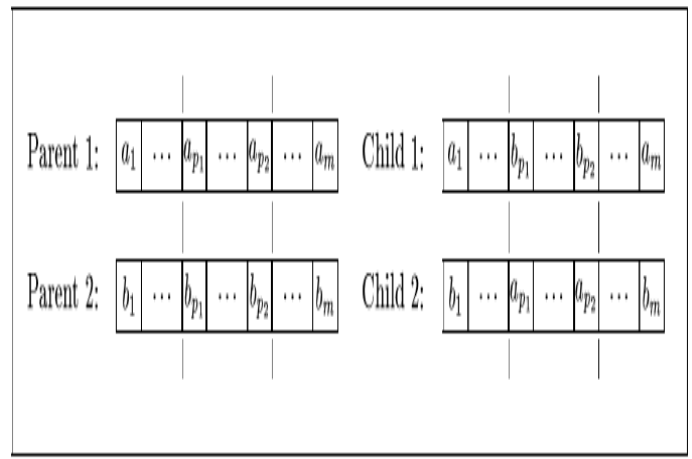

Figure 1. Two individuals (parents) are mated to generate two new individuals (children).

The mutation operator can be regarded as the mechanism that all individuals in the population are checked bit by bit and the bit values are randomly reserved according to a specified rate. Unlike crossover, this is a monadic operation. That is, a child string is produced from a single parent string. The mutation operator forces the algorithm to search new areas. Eventually, it helps the GA avoid premature convergence and find global optimal solution. The flow chart of a simple Genetic algorithm is given in Figure 2.

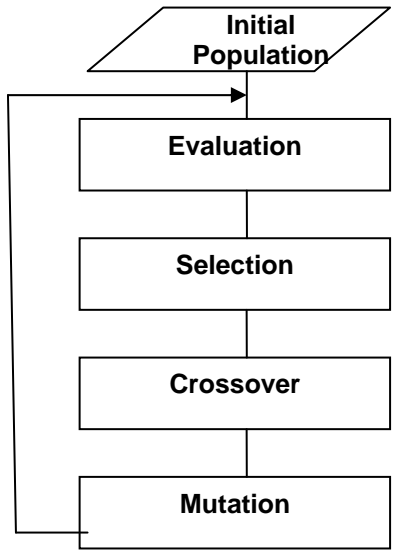

Figure 2 : Simple Genetic algorithm

*B. Tabu Search*

  *a.* The general framework Tabu Search

The basic concept of Tabu Search as described by Glover[2-3] in 1989 for solving combinatorial optimization problem. It is kind of iterative search and is characterized by the use of a flexible memory. It is able to eliminate local minima and to search beyond the local minimum. Therefore it has the ability to find the global minimum multimodal search space.

Tabu search starts from an initial solution, and at each step such a move to a neighboring solution is chosen to hopefully improve objective criterion value. This is close to a local improvement technique except for the fact that a move to a solution worse than the current solution may be accepted. Algorithm tries to take steps to assure that the method does not reenter a solution previously generated which serve as a way to avoid becoming trapped in local extreme. The variant of the algorithm we use accomplishes this by recency-based data structure called tabu list that contains the moves that are discouraged at the current iteration. A move remains a restricted one only for a limited number of iterations. Algorithm is not guaranteed to find optimum solution; however, experimental results show that even if not successful it does find good near-optimum solution.

  b. Outlines of the Basic Tabu Search Algorithm

Outlines of basic TS are described below.

1. Generate a random initial solution(S) and calculate its fitness(C). Record this as the best solution so far.

2. Create a list of possible moves: here a 'move' consists of swapping two randomly chosen elements(s, y) of the current key. The size of this list is a parameter of the algorithm and it is not necessarily fixed. The fitness of the solution obtained by making each of the moves are calculated.

3. Choose the best admissible candidate. Of the candidate moves, which one is not tabu and yields the best improvement in the fitness of the current solution? A move which is tabu may be accepted if it satisfies the aspiration criteria(A).

4. Update the tabu list and the best so far (if necessary).

5. Repeat step 2, 3 and 4 until a fixed number of iterations have been performed, or there has been no improvement in the best solution for a number of iterations.

The tabu list encloses the most recent moves. Its function is to prohibit certain move in a period of time to prevent cycling and avoid local optimum. This function is used to generate the new solution state from the current solution state which is assumed to be a best "move". If the new solution states improve the objective function value, it would be accepted as the new current solution state.

  c. Aspiration Level

The use of an aspiration level function (A) which depends on move (s,y) is one of the tabu search important features. If the objective value (C) of a move (s,y) is less than a pre specified aspiration level A , then the tabu status of the move may be overridden. The move is now defined as a solution-specific move, depending on both s and y. Each solution-specific move is characterized by a set of attributes. The aspiration level might be defined either for a collection of these or for a specific objective. Once a move passes the criterion then its tabu status is overridden.

  d. Strategies

The strategies are also an important element tabu search, referring to the search stage where the moves are allowed to enter the infeasible region. The search oscillates back and forth between the feasible and infeasible solution space. Thus it provides the opportunities to select paths which otherwise might not be allowed. Strategic oscillation might also be useful for sensitivity analysis by providing a range of solutions.

e. Intermediate and long term memory

The intermediate term memory function records features that are common to a set of best trial solutions during a particular period of the search. The search then continues, using these common features as a heuristics to identify the new solutions. The long term memory function diversifies the search from the current search stage by using a heuristic which is usually generated from the search. It generally works in a manner opposite to the intermediate memory by penalizing good moves rather than rewarding them. This step might achieve an escape from a local optimal solution. Both functions are used for a short number of iterations and then the search continues with the original heuristics or evaluation criteria.

*C. Simulated Annealing*

a. The general framework Simulated Annealing Algorithm

In 1983 Kirkpatrick[4] proposed an algorithm which is based on the analogy between the annealing of solids and the problem of solving combinatorial optimization problems.

Annealing is the physical process of heating up a solid and then cooling it down slowly until crystallizes. The atoms in the material have high energies at high temperatures and have more freedom to arrange themselves. As the temperature is reduced, the atomic energies decrease. A crystal with regular structure is obtained at the state where the system has minimum energy. If the cooling is carried out very quickly, which is known as rapid quenching, widespread irregularities and defects are seen in the crystal structure. The system does not reach the minimum energy state and ends in a polycrystalline state which has a higher energy.

In 1953 Metropolis et al. [5], showed that the distribution of energy in molecules at the "minimum energy state" is governed by the Boltzman probability distribution.

$$P(E) = e^{\left(\frac{-\Delta E}{kT}\right)} \quad (1)$$

where $\Delta E = E2 - E1$, k is Boltzmann's constant and T is the temperature. The Metropolis Algorithm uses Equation (1) to make a decision as to whether or not a transition between different energy levels will be accepted. The Metropolis Algorithm can be summarized by Equation (2),

$$P(E) = \begin{cases} 1 & \text{if } \Delta E \geq 0 \\ e^{\frac{-\Delta E}{T}} & \text{if } \Delta E < 0 \end{cases} \quad (2)$$

Consider, now, if the evaluation of the cost function for the problem being solved is equivalent to the energy in Equation 2. A transition which decreases the cost (an increase in energy) will always be accepted. However, the Metropolis Algorithm is structured so that a transition to a solution with a higher cost (lower energy) is accepted, with a probability that decreases as the temperature increases. This gives the algorithm the ability to move away from regions of local minima. This is not the case for the so-called "iterative improvement" techniques which only move in the direction of decreasing cost. That is, a transition is only accepted if $\Delta E > 0$.

In the analogy between a combinatorial optimization problem and the annealing process, the states of the solid represent feasible solutions of the optimization problem, the energies of the states correspond to the values of the objective function computed at those solutions, the minimum energy state corresponds to the optimal solution to the problem and rapid quenching can be viewed as local optimization.

b. Outlines of the Basic Simulated Annealing

The simulated annealing algorithm consists of a number of components.

- There first must exist some measure for evaluating the "goodness" of a particular configuration (or solution). This is called the cost function or the objective function.

- A cooling schedule must be determined. A cooling schedule defines the initial temperature, the way in which the temperature decreases at each iteration, and when the annealing should cease. Many complex mathematical models have been devised in consideration of the cooling schedule, however a simple model will usually suffice.

- There must be a set of rules which state how a particular solution is changed in the search for a better solution.

Outlines of basic Simulated Annealing are described below.

1. Generate an initial solution to the problem (usually random).

2. Calculate the cost of the initial solution.

3. Set the initial temperature $T = T^{(0)}$.

4. For temperature, T, do many times:

- Generate a new solution - this involves modifying the current solution in some manner.

- Calculate the cost of the modified solution.

- Determine the difference in cost between the current solution and the proposed solution.

- Consult the Metropolis Algorithm to decide if the proposed solution should be accepted.

- If the proposed solution is accepted, the required changes are made to the current solution.

5. If the stopping criterion is satisfied the algorithm ceases with the current solution, otherwise the temperature is decreased and the algorithm returns to Step 4.

   c.   The Cooling Schedule

As mentioned above, the cooling schedule has three main purposes.

1. It defines the initial temperature. This temperature is chosen to be high enough so that all proposed transitions are accepted by the Metropolis Algorithm.

2. The cooling schedule also describes how the temperature is reduced. Although there are other methods, two possibilities are presented here.

   (a)   An exponential decay in the temperature

   $$T^{(k+1)} = \alpha \times T^{(k)} = \alpha^{(k)} \times T^{(0)},$$

   where $0 < \alpha < 1$

   Usually $\alpha \approx 0.9$ but can be as high as 0:99

   (b)   Linear decay: here the overall temperature range is divided into a number of intervals, say K.

   $$T^{(k+1)} = \frac{K-k}{K} \times T^{(0)}$$, where k=1,……,K

3. Finally, the cooling schedule indicates when the annealing process should stop. This is usually referred to as the stopping criterion. In the case where a linear decay is used the algorithm can be run for its K iterations, provided K is not too large. For the case where exponential decay is used, the process usually ceases when the number of accepted transitions at a particular temperature is very small $(\approx 0)$.

III. CONCLUSION

This paper has investigated the robust and efficient methodology of three well-studied Evolutionary Computation algorithms like genetic algorithm, tabu search & simulated annealing in the field of cryptanalysis.

IV. REFERENCES

[1]   Holland J.. "Adaptation in Natural and Artificial Systems" University of Michigan Press, Ann Arbor, Michigan, 1975.

[2]   Glover F., "Tabu search ", ORSA Journal on computing, Vol.1 ,pp. 190-206, 1989

[3]   Glover Fred. "Tabu search part II", ORSA Journal on computing, Vol 2. ,pp. 14-32,1990

[4]   Kirkpatrick S., Gelatt C. D., Jr., and Vecchi M. P., "Optimization by simulated annealing", Science, 220(4598):671–680, 1983.

[5]   Metropolis N., Rosenblunth A.W., Rosenblunth M. N., Teller A.H.,1 and Teller E., "Equations of state calculations by fast computing machines". Journal of Chemical, Physics, 21(6):1087–1092, 1953.